\documentclass[%
 aip,
 amsmath,amssymb,
 groupedaddress,
 reprint,%
]{revtex4-1}

\usepackage{graphicx}
\usepackage{dcolumn}
\usepackage{bm}

\usepackage[utf8]{inputenc}
\usepackage[T1]{fontenc}
\usepackage{mathptmx}
\usepackage{textcomp} 
\usepackage{nicefrac}
\usepackage{xcolor}

\begin{document}

\preprint{AIP/123-QED}

\title[Time-Resolved Momentum Microscopy with a 1 MHz High-Harmonic Extreme Ultraviolet Beamline]{Time-Resolved Momentum Microscopy with a 1 MHz High-Harmonic Extreme Ultraviolet Beamline}


\author{Marius Keunecke} \email{mkeunec@gwdg.de}%
\author{Christina Möller}%
\author{David Schmitt} %
\author{Hendrik Nolte} %
\author{G. S. Matthijs Jansen} %
\author{Marcel Reutzel} %
\author{Marie Gutberlet} %
\author{Gyula Halasi} \thanks{Current address: ELI-ALPS, ELI-HU Non-Profit Ltd., H-6720 Szeged, Dugonics tér 13, Hungary}%
\author{Daniel Steil} %
\author{Sabine Steil} 
\author{Stefan Mathias} \email{smathias@uni-goettingen.de}%

\affiliation{I. Physikalisches Institut, Georg-August-Universit\"at G\"ottingen, Friedrich-Hund-Platz 1, 37077 G\"ottingen, Germany}

\begin{abstract}
Recent progress in laser-based high-repetition rate extreme ultraviolet (EUV) lightsources and multidimensional photoelectron spectroscopy enable the build-up of a new generation of time-resolved photoemission experiments. Here, we present a setup for time-resolved momentum microscopy driven by a 1~MHz femtosecond EUV table-top light source optimized for the generation of 26.5~eV photons. The setup provides simultaneous access to the temporal evolution of the photoelectron´s kinetic energy and in-plane momentum. We discuss opportunities and limitations of our new experiment based on a series of static and time-resolved measurements on graphene. 
\end{abstract}

\maketitle


\section{\label{sec:intro}INTRODUCTION}

\begin{figure*}[hbt!]
    \centering
    \includegraphics[width=\textwidth]{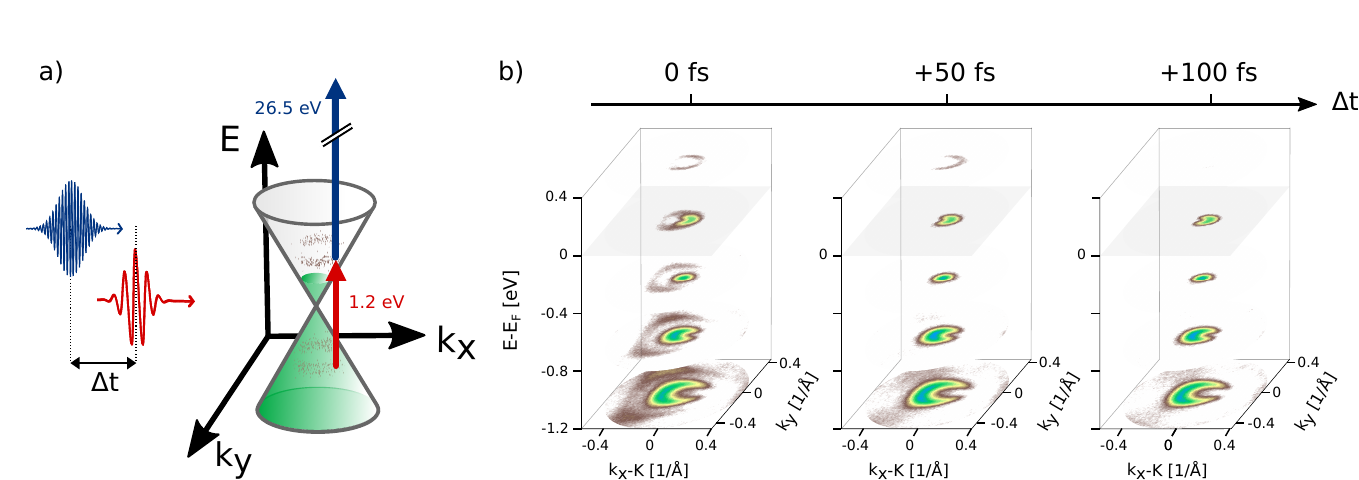}
    \caption{Energy- and in-plane-momentum resolved photoemission data showing the electron dynamics at the K-point of graphene upon illumination by an infrared femtosecond laser pulse. (a) Resonant driving with IR photons (red arrow) promotes electrons into the former unoccupied part of the Dirac cone of n-doped graphene; the hot charge carrier distribution is probed by the EUV pulse (blue arrow). (b) Three-dimensional data stacks illustrating the photoemission yield (color-coded) as a function of energy and in-plane momentum. From left to right, the pump-probe delay $\Delta t$ is increased; the gray-shaded plane indicates the Fermi energy.  In temporal overlap, photoemission side-bands and hot charge carriers above the Fermi energy are clearly resolved. With increasing delay, the side-bands vanish and the hot charge-carrier distribution thermalizes.}
    \label{fig:Time-resolved_band_structure}
\end{figure*}

Full spectroscopic information on the electronic band structure of a solid-state material requires the measurement of multiple observables. In a photoemission experiment, the photoelectron's energy, momentum, and spin are desirable quantities.\cite{hufner_photolectron_2003} Novel surface science tools, termed momentum microscopes (MM), can provide simultaneous access to these quantities.\cite{kromker_development_2008, Kolbe:2011eb, schonhense_space-_2015, tusche_spin_2015} In first experiments, MM has been applied to band mapping of various material systems and complex adsorbate structures.\cite{tusche_multi-mhz_2016, medjanik_direct_2017, fedchenko_high-resolution_2019, xian_open-source_2019, jansen_efficient_2020} In selected examples, it has been shown that the combination of time-of-flight momentum microscopy with femtosecond pump-probe spectroscopy enables the collection of multidimensional data sets describing the ultrafast non-equilibrium dynamics of the material under strong illumination. \cite{haag_time-resolved_2019, kutnyakhov_time-_2020} An example for the ultrafast evolution of such three-dimensional data stacks containing energy, and in-plane-momentum information is shown in Fig.~\ref{fig:Time-resolved_band_structure}.


Time-resolved momentum microscopy (TR-MM) has recently been demonstrated as an impressive tool to map hot charge-carrier dynamics in momentum space after excitation by the frequency-doubled output of a titanium-sapphire laser operating at 80~MHz.\cite{haag_time-resolved_2019} However, the accessible range of in-plane momenta, $k_{x,y}$, is fairly limited 
in this experiment due to the use of light pulses in the visible range of the spectrum. The accessible momentum range is given by
\begin{equation}
\begin{split}
k_{x,y}\left(\mathring{A}^{-1}\right) & = 0.514\sqrt{E_{\rm kin}\left({\rm eV}\right)} \\ 
& = 0.514\sqrt{\hbar\omega\left({\rm eV}\right)-\phi\left({\rm eV}\right)-E_{\rm B}\left({\rm eV}\right)},
\label{eq:photoemission_horizon}
\end{split}
\end{equation}
in which $E_{\rm kin}$ is the photoelectron kinetic energy, $\hbar \omega$ the  photon energy, $\phi$ the material work function and $E_{\rm B}$ the binding energy.\cite{hufner_photolectron_2003} In order to explore the material's whole Brillouin zone, the kinetic energy of the photoelectrons needs to be increased, for example through the use of extreme ultraviolet (EUV) photons with energies between 10 and 124~eV.\cite{Bauer:2005cp, mathias_angle-resolved_2007} For TR-MM, free electron lasers (FELs) and high-harmonic generation (HHG) are promising EUV light sources, which offer ultrashort femtosecond or even attosecond EUV pulses.\cite{chang97, Hentschel:2001wx, Spielmann97, krausz_attosecond_2009, Popmintchev:2012gh, harmand_achieving_2013, duris_tunable_2020} 

In photoemission experiments with a pulsed excitation source, a high repetition rate of the driving laser is desired to maximize counting statistics. Early time-resolved optical-pump---EUV-probe photoemission experiments, however, have been limited by the low (kilohertz) repetition rate of available laser amplifier systems.\cite{Bauer:2005cp, Haarlammert:2009cw, Rohwer:2011fy, Cavalieri:2007de, Schultze:2010kr, Petersen:2011kt, Gierz13natmat, Johannsen13prl, Tao:2016in, Mathias16natcom, Eich:2017im, Stadtmuller:2019fe, Siek:2017cq, Wallauer16apa, sie_time-resolved_2019, Frietsch:2015dr, LiuXYZ2019} Here, the large number of photoemitted electrons per pulse could lead to signal distortions due to space charge effects.\cite{passlack_space_2006, Hellmann09prb, Oloff16jap, Schonhense18njp} Avoiding these has so far hindered satisfactory counting statistics at manageable measurement duration. 
Recent developments of femtosecond laser technology and table-top HHG sources have made EUV sources operating at MHz-level repetition rates widely available,\cite{Heyl12jpb, hadrich_high_2014, carstens_high-harmonic_2016, harth_compact_2017} which are particularly promising for time-resolved photoemission spectroscopy.\cite{chiang_boosting_2015, puppin_time-_2019, corder_ultrafast_2018, saule_high-flux_2019} 

In this article, we present the powerful combination of a table-top HHG beamline driven by a fiber laser system, and a flexible pump beamline, all operating at the exceptional repetition rate of 1~MHz, with a momentum microscope endstation guaranteeing simultaneous detection of energy- and in-plane-momentum-resolved photoemission data. In a time-resolved optical pump - EUV probe configuration, the femtosecond time evolution of non-equilibrium charge carrier and band renormalization dynamics can be followed with unprecedented information depth in the full Brillouin zone. Here, we will demonstrate the capabilities of this novel system based on exemplary measurements on monolayer graphene. 

\section{\label{sec:EUV}The femtosecond 1~MHz EUV beamline}

An ideal light source for time-resolved momentum microscopy has to fulfill several key requirements with respect to the photon energy, intensity, repetition rate, pulse width and spectral bandwidth. The photon energy should be at least in the EUV regime in order to access the full Brillouin zone of the electronic band structure (see equation 1). For a maximal detection rate of a time-of-flight detector with minimal distortions due to space-charge effects, a MHz repetition rate light source with a moderate number of photons per pulse is desirable. Also, time-resolution and spectral bandwidth, which are connected via the time-bandwidth product, need to be adapted to the envisaged experiments.  The light source should also provide synchronized pump pulses with tunable wavelength, pulse duration and intensity. In this section, we describe such a beamline based on a table-top, high repetition-rate fiber amplifier laser.

\subsection{High harmonic generation and EUV monochromatization \label{sec:euv_beamline}}

\begin{figure*}[htb]
\includegraphics[width=\textwidth]{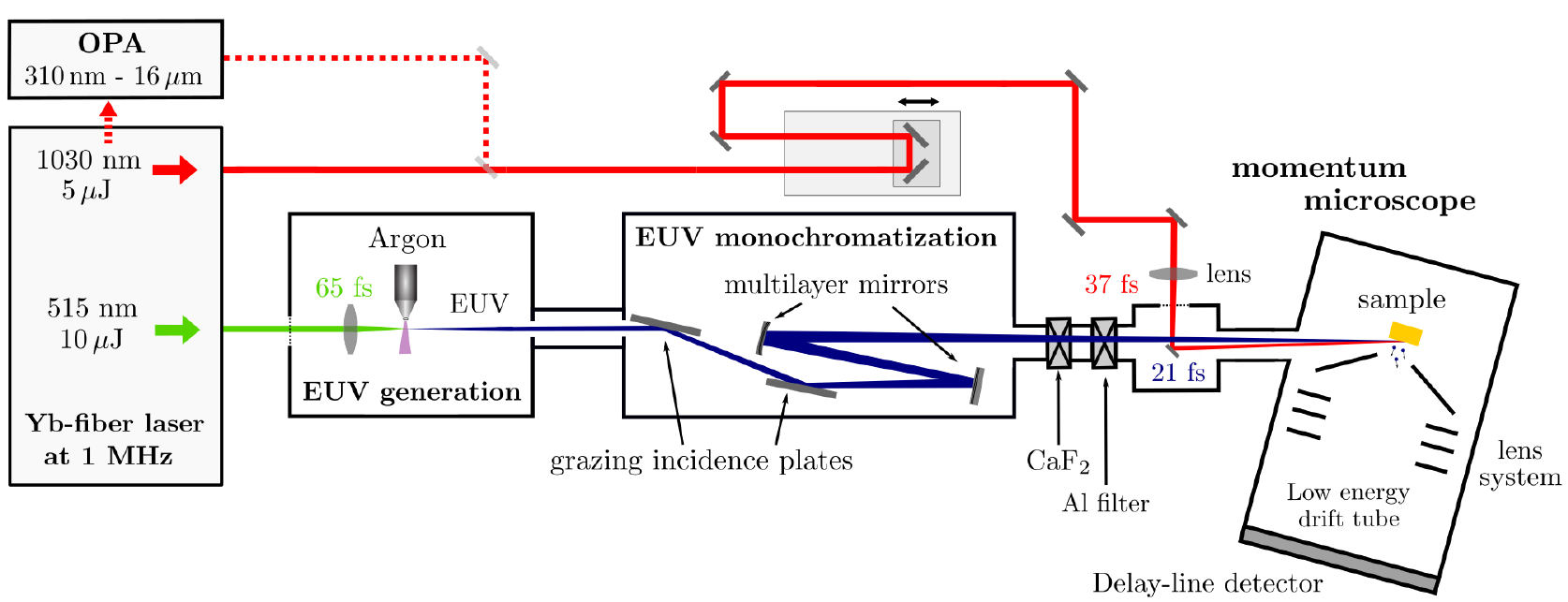}
\caption{Schematic layout of the experimental setup consisting of the 1 MHz EUV beamline, the pump line and the momentum microscope. A detailed description is given in Sec.~\ref{sec:EUV} and ~\ref{sec:endstation}. \label{fig:setup}}
\end{figure*}

Figure~\ref{fig:setup} shows a scheme of the experimental realization of the EUV light source together with the photoemission endstation. We use an ytterbium-doped fiber amplifier (Active Fiber Systems) which delivers 100~\textmu J pulses at a repetition rate of 1~MHz with a pulse length of approximately 250-300~fs. These pulses are spectrally broadened through self-phase modulation in a 1~m hollow-core fiber filled with 12~bar krypton gas and compressed with a pair of chirped mirrors resulting in a pulse duration of $\sim$35-40~fs at 50~\textmu J pulse energy.\cite{nisoli_compression_1997, rothhardt_53w_2014} 

The compressed pulses are frequency-doubled to 515~nm (10~\textmu J pulse energy) before being focused in an argon gas jet to generate high harmonics.\cite{Eich14jesrp} Considering the high repetition rate of 1~MHz and the resulting low pulse energies, we generate HHG in the tight-focusing regime with a 75~mm focal length lens.\cite{Heyl12jpb, Chiang2012} The resulting focal spot measures 12~\textmu m full-width at half maximum. The pulse duration is measured by intensity autocorrelation to be 65~fs (Gaussian), mainly due to dispersion in the thin lens and the entrance window. This corresponds to a peak intensity of $5\cdot 10^{13}$~W~cm\textsuperscript{-2} in the focus. The gas jet is mounted on a high precision 3-axes position system to optimize the gas jet position relative to the laser focus. 
We observe efficient HHG at an argon backing pressure of 2~bar for a gas nozzle diameter of 100~\textmu m. The gas nozzle is placed 150~\textmu m behind the laser focus where a dominant contribution from the short HHG trajectory is expected.\cite{Heyl12jpb} The acceptance angle of the systems corresponds to a full-angle divergence of 20 mrad whereas the EUV divergence is slightly higher. In order to limit reabsorption of the generated EUV light, we pump the system with turbomolecular pumps backed by a multi-stage root pump. The resulting pressure with argon gas load in the EUV generation and monochromatization chambers, which are separated by a differential pumping stage, is $2\cdot 10^{-3}$~mbar and $5\cdot10^{-5}$~mbar, respectively. 

\begin{figure}[htb]
\centering
\includegraphics[width=\linewidth]{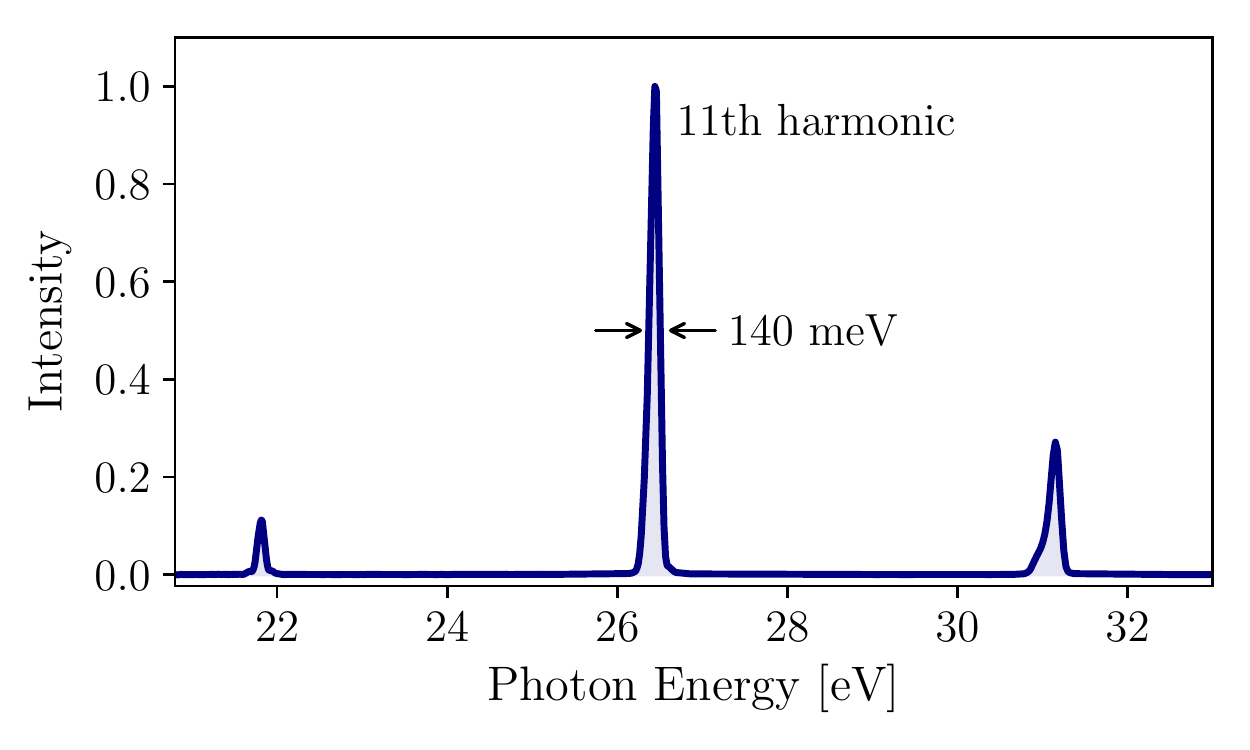}
\caption{Normalized high harmonic spectrum generated in an argon gas jet with a 10~\textmu J, 515~nm driver operating at 1~MHz. The 11\textsuperscript{th} harmonic at 26.5~eV with a bandwidth of 140~meV is used for photoemission spectroscopy in the MM.
\label{fig:HHG_spectrum}}
\end{figure}

The HHG spectrum shown in Fig.~\ref{fig:HHG_spectrum} is measured by placing a gold mirror into the beam path just before the multilayer mirrors. Our home-built spectrometer consists of a toroidal mirror, a 500~lines/mm grating mounted in the off-axis geometry\cite{Neviere1978} and an EUV CCD camera. We generate harmonics between 20 and 32~eV, separated by 4.8~eV -- twice the photon energy of the HHG pump beam.
The 11\textsuperscript{th} harmonic at 26.5~eV has a bandwidth of 140~meV for the above specified parameters of the HHG pump pulses,  which corresponds to a relative bandwidth of $\nicefrac{\Delta E}{E}=5.3\cdot 10^{-3}$. 
Correcting for the CCD efficiency, spectrometer and filter transmission, we estimate an average generated power of 11.5~\textmu W for the 11\textsuperscript{th} harmonic. This corresponds to $2.7\cdot 10^{12}$ generated photons per second and a conversion efficiency of $1.5\cdot 10^{-6}$.
 
The EUV light is separated from the HHG driver by a pair of so-called grazing incidence plates (GIP),\cite{pronin_ultrabroadband_2011} which consist of a fused-silica substrate with an anti-reflection coating for 515~nm and a top-layer of SiO$_{2}$. This allows for high reflection of the high harmonics (50\% for 26.5~eV) at a grazing incidence angle of 10\textdegree{} with simultaneous high transmission for the HHG driver. In order to avoid overlapping photoemission spectra excited by neighboring high harmonics, the selection of one single harmonic is necessary. This is achieved using a double mirror monochromator with two multilayer mirrors (optiX fab GmbH), which reflect the 11\textsuperscript{th} harmonic around 46.6~nm in a 5~nm bandwidth. This configuration preserves the pulse duration, in contrast to single-grating monochromatization schemes which suffer from spatial chirp. The monochromator has a total transmission of 9\% at 5\textdegree{} angle of incidence on the mirrors. We measured an extinction ratio of $\sim$1:470 with respect to the 13\textsuperscript{th} harmonic. 

We note that the reflectance and thereby also the monochromatization properties of these mirrors are altered by carbon contamination, which is a well-known problem for EUV and X-ray beamlines.\cite{Hollenshead2006, Koide1988} Regular ozone cleaning with the help of an UV lamp avoids the reduction of the EUV flux.\cite{Hansen1993}
The last multilayer mirror has a radius of curvature of 1200~mm and is positioned such that the EUV beam is focused onto the sample in the momentum microscope. A 200~nm thick, free-standing aluminum foil is used to block residual light from the fundamental beam and has a measured transmission of $12\,\%$ for the 11\textsuperscript{th} harmonic, limited by thin oxidation layers on the aluminum surface. The aluminum foil is mounted in a vacuum valve, thereby separating the ultrahigh, $< 5\cdot10^{-10}$~mbar vacuum in the momentum microscope from the high vacuum in the preceding chambers. 
In total, we expect 0.3\% of the generated 26.5~eV photons to reach the sample in the momentum microscope resulting in $8.5\cdot 10^3$ photons per pulse at the sample. We note that other schemes of the monochromatization, e.g. with metal filters only,\cite{Eich14jesrp} can be considerably more efficient. However, the generated flux with our beamline is more than sufficient to photoemit > 1 electron/pulse, which corresponds to the limit of the detection rate of our time-of-flight-based momentum microscope. 

\subsection{A versatile pump beamline \label{sec:pump_beamline}}

A versatile pump beamline with tunable wavelength, pulse duration, polarization, and intensity enables pump-probe photoelectron spectroscopy on a wide range of sample systems, addressing, for example, resonant transitions in many different materials. We derive the pump pulses from the fiber laser system using a 90:10 beam splitter, positioned after the nonlinear pulse compression. The $90\%$ output is used for HHG as described above in section~\ref{sec:euv_beamline}. The residual $10\%$ is guided around the EUV generation and monochromatization chambers onto a delay stage to control the relative pump-probe timing. An attenuator and a $\lambda /2$ waveplate in the pump beamline allow for manipulation of the polarization and beam intensity. Different pump wavelengths can be achieved through various nonlinear frequency conversion stages that can be inserted in the beamline. This provides intense, femtosecond pump pulses at the fundamental wavelength (1030~nm) and its higher harmonics (515~nm, 343~nm, 258~nm). Furthermore, a part of the uncompressed, 300~fs laser pulses can be split off to pump an optical parametric amplifier (OPA, Orpheus-F/HP from Light Conversion). This enables us to excite at wavelengths ranging from  310~nm up to 16~\textmu m. We note that in the current setup, for efficient usage of the OPA, the fiber laser has to be operated at 500~kHz. 
After the aluminium filter, the pump beam is coupled into the MM by a mirror positioned such that the angle of the pump beam with respect to the EUV beamline is less than 1\textdegree{} (Fig.~\ref{fig:setup}). This prevents loss of temporal resolution due to non-collinear pump and probe beams, and ensures optimal time-resolution limited only by the individual pulse durations.

\section{\label{sec:mm_general}The photoelectron momentum microscope}

In this section, we provide a general description of the surface science endstation. Subsequently, we characterize the capabilities of the beamline for static band mapping, as well as real-time monitoring of photo-induced non-equilibrium dynamics.

\subsection{\label{sec:endstation}Surface Science endstation}

The outstanding potential of time-of-flight momentum microscopy lies in the simultaneous detection of three-dimensional data sets containing energy- and in-plane-momentum-resolved information on the detected photoelectrons. The general working principle of our time-of-flight momentum microscope (METIS, Surface Concept GmbH) is described in Ref.~\onlinecite{medjanik_direct_2017}. In short, electrons emitted from the surface region are collected by an electrostatic lens (extractor) with a voltage of up to 29~kV. This enables the collection of photoelectrons with emission angles of up to $\pm 90^\circ$ for photoelectron kinetic energies up to 70~eV, maximizing the accessible in-plane momentum range in a single measurement without the need to rotate the sample. Subsequently, a compound lens system maps the momentum-resolved photoelectron distribution obtained in the back focal plane of the objective lens through a time-of-flight drift tube onto a position- and time-sensitive delay-line-detector (DLD4040 R2.55, Surface Concept). While the in-plane momentum information is obtained by the two-dimensional point of impact, the energy information is measured by the moment of incidence onto the detector. \cite{Oelsner2001} Thus, by limiting the detection to one photoelectron per excitation light pulse, for each individual detected photoelectron full ($E,k_{x},k_{y}$)-information can be stored into a 3D histogram. Such a complete measurement allows for extensive post-processing, such as the correction of timing jitter, pulse energy, or other pulse-to-pulse variations. However, we note that in contrast to measurements at FELs, such major post-processing is not necessary in our case due to the stability of the light source. A useful example for the calibration and post-processing of the raw measurement data is summarized in Ref.~\onlinecite{xian_open-source_2019}. 

The specimen is mounted onto a motorized hexapod for alignment with respect to the electrostatic lens system of the microscope. A continuous-flow cryostat is integrated into the hexapod to facilitate cooling of the sample to below 20~K. 
The EUV and pump beams are directed onto the sample at a grazing incidence angle of 22$^\circ$.

Multiple samples can be stored in a separate preparation chamber that is equipped with a wide range of surface science tools for specimen preparation and characterization, including sputtering and annealing, evaporators, and low energy electron diffraction. The preparation chamber is also equipped with a load lock for fast sample exchange. 

\subsection{\label{sec:static}Static Momentum Microscopy}

The EUV beamline combined with the MM is a versatile tool for mapping the equilibrium band structure of solid state materials as it enables the simultaneous measurement of ($E$, $k_x$, $k_y$)-resolved photoelectron distributions. 
Very high energy- and momentum-resolution better than <15~meV and <0.01~\AA\textsuperscript{-1} have already been achieved using a hemisphere-based momentum microscope in combination with a helium-discharge light source\cite{tusche_spin_2015} (note that the same electrostatic lens system is used in our setup in combination with a time-of-flight detector). In this work, the momentum and energy resolution is limited by the bandwidth of the photoemission light source, the precise settings of the electrostatic optics, and the used drift energy. In our case, we will provide an estimate of the energy-resolution of our setup in operation at the end of this section.

\begin{figure}[htb!]
    \centering
    \includegraphics[width=\linewidth]{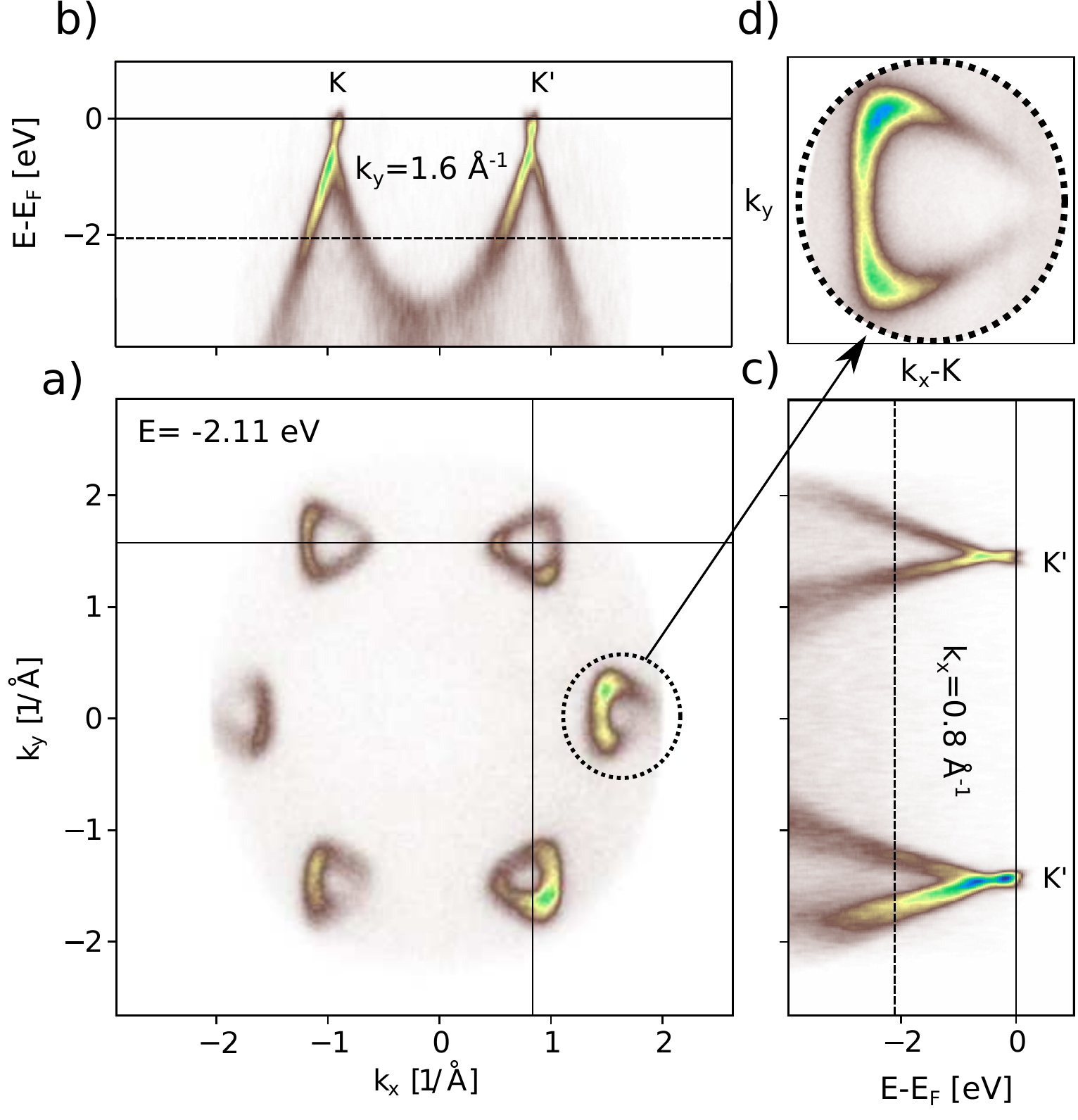}
    \caption{Static photoemission spectroscopy at room temperature with 26.5~eV photons showing the band structure of $n$-doped graphene. (a) Photoelectron momentum distribution at  $E-E_F =$2.1~eV. (b,c) Selected cuts through the 3D data in ($E,k_x$) and ($E,k_y$) direction. (d) Measurement with higher MM magnification settings and focus on a single Dirac Cone, showing the dark corridor to the right.}
    \label{fig:Occupied_Band_structure_Graphene}
\end{figure}

We performed exemplary measurements on an n-doped graphene monolayer on a 4H-SiC (0001) substrate (10~mm x 5~mm), which was prepared by the polymer-assisted sublimation growth technique\cite{kruskopf2016comeback, momeni2018minimum} and cleaned by subsequent heating cycles up to 600 K in ultra-high vacuum. Fig.~\ref{fig:Occupied_Band_structure_Graphene} shows band mapping data from photoemission with 26.5~eV photons to illustrate the overall performance of the setup. Satisfying signal quality within the full ($E$, $k_x$, $k_y$)-stack is achieved in an integration time of 15~min when the experiment is carried out at 1~MHz repetition rate in $p$-polarized excitation geometry. The bottom-left plot shows a ($k_x$, $k_y$)-resolved energy slice taken at $E-E_F =$2.1~eV; the first Brillouin zone of graphene can be identified by the observation of six Dirac cones at the K and K' points, for which the so-called dark corridor is clearly resolved.\cite{Bostwick07natphys, Gierz:2011do} From the ($E$, $k_x$, $k_y$)-resolved data stacks, arbitrary cuts in the ($E$, $k_x$)- and ($E$, $k_y$)-directions can be extracted; Fig.~\ref{fig:Occupied_Band_structure_Graphene}~(b,c) shows such cuts along the K-K' and K'-K' directions. 
The data in Fig.~\ref{fig:Occupied_Band_structure_Graphene} has been corrected for  distortions arising from the electrostatic lens system. This is achieved using a set of stretching and shearing matrices which optimize the known hexagonal symmetry of the Dirac points.\cite{xian_open-source_2019} The plane of incidence of the p-polarized EUV beam is aligned along the horizontal axis of the microscope, but 10\textdegree{} off with respect to the $\rm \Gamma$-K direction of the graphene sample; this twist between detector-coordinates and $\rm \Gamma$-K is post-corrected in the data shown in Fig.~\ref{fig:Occupied_Band_structure_Graphene}. The distortion-corrected band structure agrees well with the band structure predicted by a tight-binding calculation.\cite{CastroNeto2009} 
The presented analysis highlights that the MM in operation with the table-top pulsed EUV beamline is well suited for band mapping.

For MM studies that concentrate on smaller in-plane momentum regions, counting statistics can be significantly improved by changing the electrostatic lens system such that only the desired region of interest is mapped onto the detector. This allows for an increase of EUV flux while staying below the detection limit of the delayline detector, effectively measuring more photoelectrons in a reduced region of interest. Fig.~\ref{fig:Occupied_Band_structure_Graphene}~(d) shows exemplary data for such a scenario, where the right-most K point is mapped in a $\sim$ 1~x~1~$\mathring{A}^{-2}$ window. 

From our measurements, we estimate the energy resolution of the setup by fitting photoelectron spectra integrated over the full measured in-plane momentum with a Fermi-Dirac distribution convolved with a Gaussian function. In our daily measurement routine, we extract a FWHM of $200\pm30$~meV for the Gaussian broadening. This broadening is composed of the linewidth of the EUV light (Fig.~\ref{fig:HHG_spectrum}) as well as the energy resolution of the MM, which is mainly determined by the used drift energy in the TOF.

\subsection{\label{sec:trmm} Time-resolved momentum microscopy}

The experiment is designed for the study of ultrafast non-equilibrium dynamics in the whole Brillouin zone. We demonstrate the performance of this combined setup using graphene as a model system. We provide a brief description of the alignment of the optical-pump-EUV probe experiment, and characterize the temporal resolution of the setup.

Initial alignment of the system is performed in a microscope setting where the real-space image of the sample is projected onto the detection plane. In such an experiment, the effective beam diameters of the 1030~nm pump and the EUV probe pulses are determined to be 100~\textmu m x 230~\textmu m and 600~\textmu m x 900~\textmu m, respectively. The elliptical shapes arise due to oblique angle-of-incidence of the beams combined with astigmatism in the EUV beamline. The real-space mapping then allows the convenient optimization of the spatial overlap of the pump and EUV beams. Within the microscope, furthermore, a field aperture can be placed into the real-space image to isolate a region where spatial overlap of the pump and probe beams is guaranteed. Temporal overlap of the infrared pump and EUV probe beams can be optimized in a two-step process: First, while scanning the delay between the pump and the HHG driver beam, the multi-photon photoemission yield is monitored and optimized for maximum count rate. Second, temporal overlap between the pump and the EUV light is verified by optimizing the photoelectron count rate in the unoccupied states above $E_{F}$.

In the following, we present exemplary femtosecond TR-MM data obtained on graphene. The hot charge-carrier dynamics of graphene, as well as of its bulk analogue graphite, have been studied in detail using time- and angle-resolved photoelectron spectroscopy (TR-ARPES).\cite{Niesner12prb, Armbrust12prl, Johannsen13prl, Gierz13natmat, aeschlimann_ultrafast_2017, Tan17prx, Johannsen:2015ds, Rode18prl} Typically, these experiments have been performed with hemispherical electron analyzers mapping energy- and $k_x$- or $k_y$-resolved photoelectron distributions onto two-dimensional imaging detectors. Subsequent scanning of the samples azimuth angle provides access to ($k_x$, $k_y$)-resolved photoemission data (e.g. Ref.~\onlinecite{aeschlimann_ultrafast_2017}). 

\begin{figure*}[htb]
    \includegraphics*[width=0.8\textwidth]{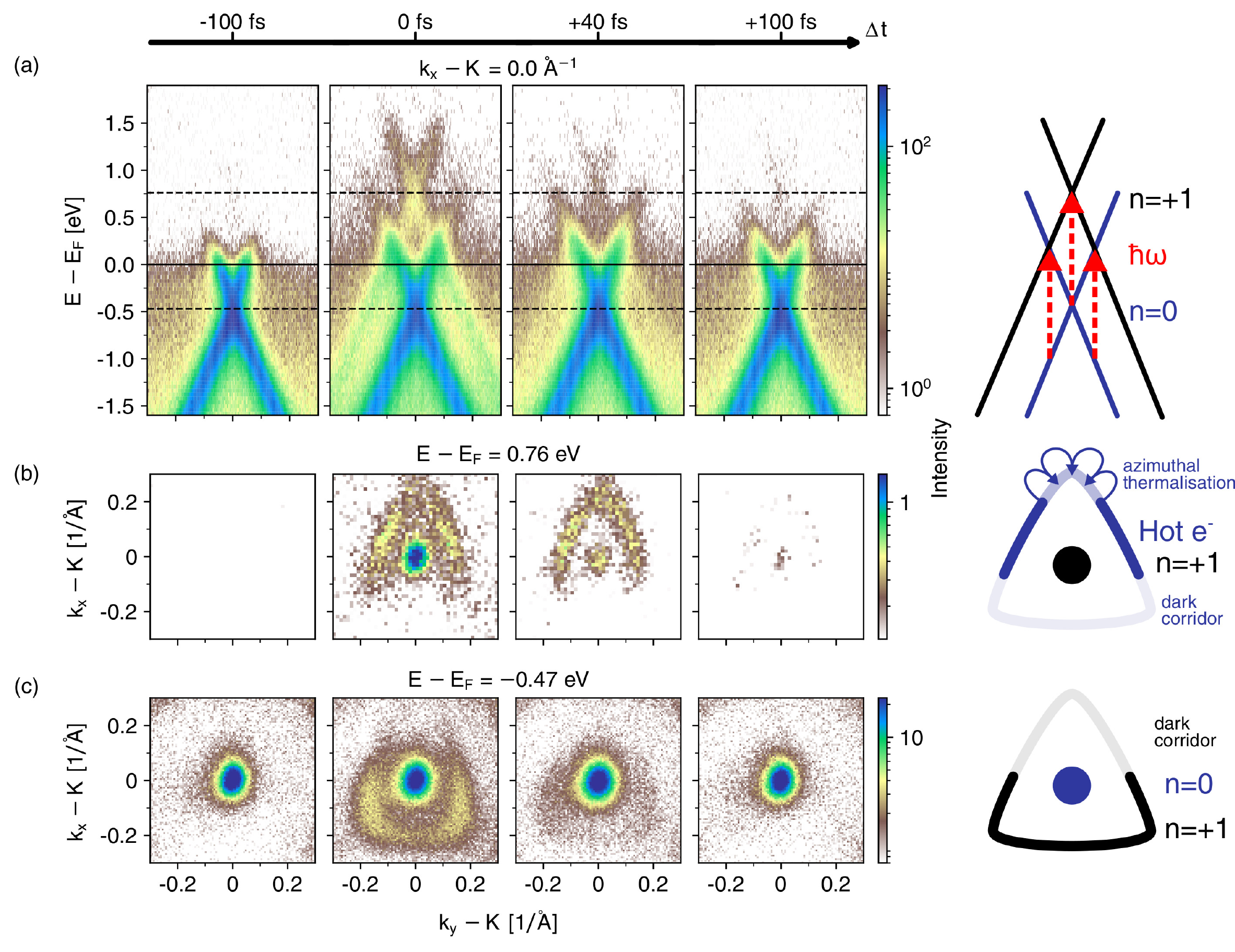}
    \caption{Time-resolved momentum microscopy of graphene. (a) The ARPES spectra are obtained by slicing the 3D data cube at the K point in $k_y$ direction and integrating over 0.03~\AA\textsuperscript{-1} along $k_{x}$. The two dotted lines indicate the position of the momentum cuts shown in (b) and (c) respectively. At temporal overlap ($\Delta t=0$~fs) a clear signature of the sideband is seen, shifted in energy by the pump photon energy of 1.2 eV. Also, the anisotropic distribution of excited charge carriers within the dirac cone\cite{aeschlimann_ultrafast_2017, Gierz:2011do} and its relaxation dynamics is clearly visible.}
    \label{fig:trimages}
\end{figure*}

The TR-MM experiment provides simultaneous access to the non-equilibrium dynamics occurring in the energy and in-plane momentum dispersive $\pi$- and $\pi^*$-bands of graphene. We performed TR-MM in the same in-plane momentum ($k_x$, $k_y$) area as in static spectroscopy (Fig.~\ref{fig:Occupied_Band_structure_Graphene}), as illustrated in  Fig.~\ref{fig:Time-resolved_band_structure}. Fig.~\ref{fig:trimages} shows selected delay-dependent ($E$, $k_y$)- and ($k_x$, $k_y$)-cuts through the measured data. In these experiments, graphene was excited with the compressed output of the fiber laser (1.2~eV, 6.5~mJ/cm$^2$, $p$-polarized), and probed by a time-delayed EUV light pulse (26.5~eV, $p$-polarized). Two dynamical processes can be identified in the data: i) Resonant excitation from the occupied into the unoccupied part of the Dirac cone leads to a non-equilibrium distribution of hot electrons, as indicated in the excitation diagram in Fig.~\ref{fig:Time-resolved_band_structure}. The thermalization of the hot charge carriers can be followed in real-time. ii) In temporal overlap of the pump and probe pulses, side-bands spaced by one pump photon energy from the main photoemission spectral feature are resolved. In the following, we present a brief description of both processes. Thereby, we focus on the capabilities of the TR-MM experiment and point towards related references that treat the observed phenomena.

As the ultrafast thermalization of hot charge carriers in the Dirac cone of graphene has been studied in detail by several groups,\cite{Niesner12prb, Armbrust12prl, Johannsen13prl, Gierz13natmat, aeschlimann_ultrafast_2017, Tan17prx, Johannsen:2015ds, Rode18prl} we do not provide a quantitative analysis of the delay-dependent dynamics of our data at this point. However, we give a qualitative description of the multi-dimensional MM data for selected time delays to indicate the experimental capabilities offered by the setup (Fig.~\ref{fig:trimages}): In temporal overlap, photoexcitation with 1.2~eV pump pulses creates an asymmetric distribution of electrons and holes in the Dirac cone, as defined by the pseudospin of the Dirac electrons with respect to the polarization of the impinging pump light;\cite{aeschlimann_ultrafast_2017} the time-dependent thermalization of these quasiparticles and their underlying scattering processes can be followed in real-time with high energy and momentum resolution. For excitation with $p$-polarized light, enhanced population of the initially unoccupied Dirac bands is localized around $k_x-K\approx 0$~\AA\textsuperscript{-1} (Fig.~\ref{fig:trimages}~(b), $\Delta t=$0~fs). With increasing $\Delta t$, azimuthal thermalization is observed ($\Delta t=$40~fs), leading to a more homogeneous charge carrier distribution that subsequently decays by 
electron-phonon scattering.\cite{Gierz13natmat} 

Replicated photoemission spectral features separated by the pump photon energy from the main photoemission line are of particular interest in the framework of light-induced band structure engineering, also known as Floquet engineering; the equilibrium band structure of solid state materials might be interfered by the time-periodic driving field, potentially manipulating band curvatures and even band topology\cite{oka_floquet_2019}. However, the separation of Floquet-Bloch bands from laser assisted photoemission (LAPE)\cite{saathoff_laser-assisted_2008} is not trivial, as has been shown in TR-ARPES data obtained on topological insulators at the $\rm \bar{\Gamma}$-point,\cite{mahmood_selective_2016, park_interference_2014} Here, we highlight that we can resolve these side-bands at large in-plane momenta at the edges of the Brillouin zone with good ($E$, $k_x$, $k_y$)-resolution (Fig.~\ref{fig:trimages}). 
Such data facilitates the separation of Floquet and LAPE contributions to these side-bands by following the routes discussed in Refs.~\cite{park_interference_2014, mahmood_selective_2016}.

At this point, we want to note that the generation of EUV light with the second harmonic\cite{Eich14jesrp} of the amplifier system is advantageous for TR-MM. Although the neighbouring 9\textsuperscript{th} and 13\textsuperscript{th} harmonics are filtered in the EUV beamline by the multilayer mirrors, residual intensity does lead to photoelectrons that are detected in our experiment. This results in replicas of the main photoemission spectral features at higher energies, spaced by $2\times$ the driving laser frequency (not shown). For the current setup, 0.25\% of the total photoemission signal is due to the 13\textsuperscript{th} harmonic. The advantage of a higher HHG driving frequency, here the second harmonic, is therefore twofold: it reduces the bandwidth requirement for the multilayer mirrors and it guarantees a maximal energetic separation of the photoemission spectral features, facilitating background free band mapping.\cite{Eich14jesrp} Such considerations are especially important when studying side-band formation, whose comparably weak photoemission signal might be overlapped with signal from a neighboring harmonic.

\begin{figure}[htb]
    \centering
    \includegraphics[width=\linewidth]{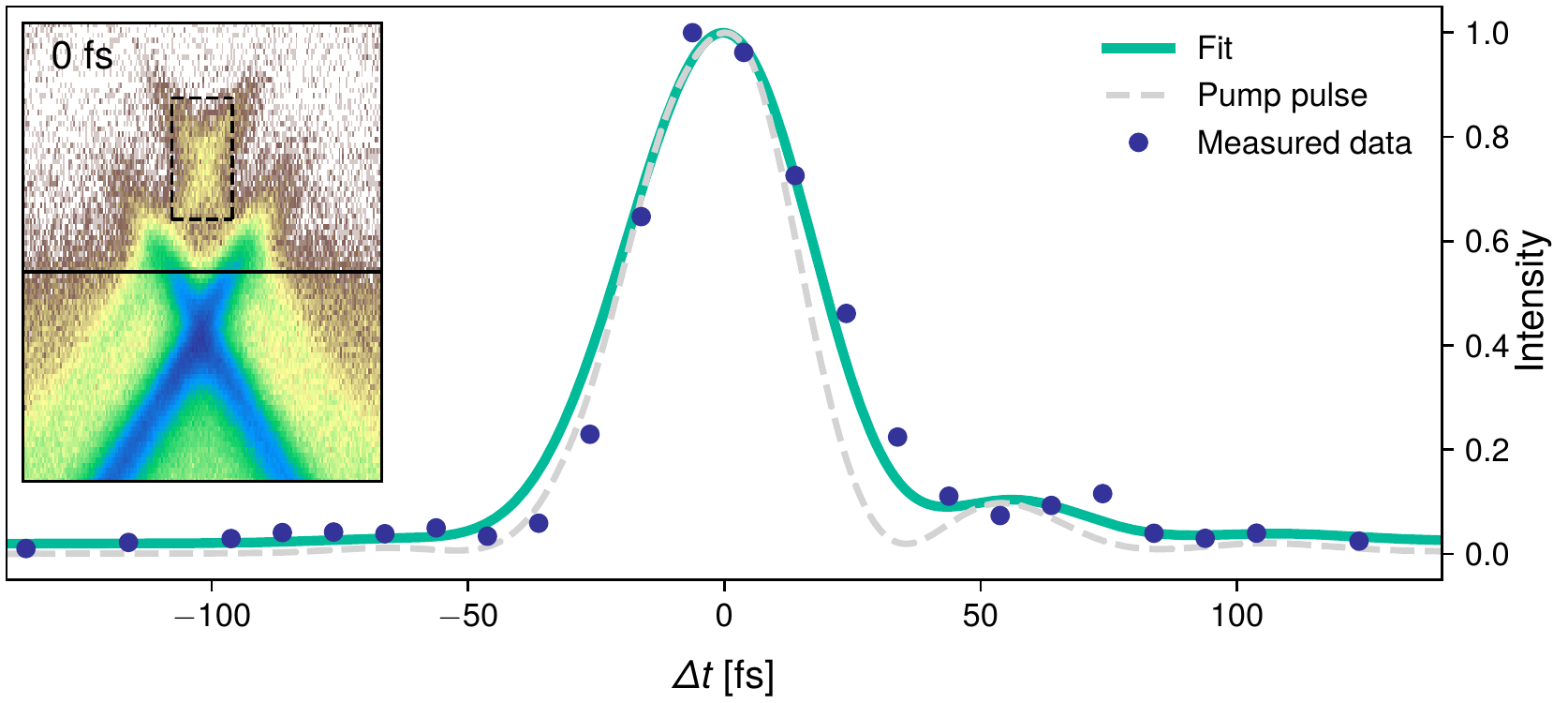}
    \caption{Time-dependence of the sideband generation (blue dots), along with the independently measured pump pulse (dashed grey line) and a fit to the data (green line), in which a Gaussian probe pulse is convolved with the pump pulse. This yields a probe pulse duration of $21\pm5$~fs.
    The inset shows the measured ARPES spectrum at $\Delta t=0$~fs. The box indicates the area over which the sideband signal is integrated. Note that we carried out an additional measurement here, where we determined and optimized the pump pulse duration using FROG.} 
    \label{fig:LAPE_trace}
\end{figure}

Finally, we make use of the sideband generation to characterize the time-resolution of the TR-MM experiment. The intensity of the sidebands, independent of their physical origin, 
follows the intensity of the pump laser pulse. Therefore, we can measure the cross-correlation between the IR pump and the EUV probe pulse by integrating the photoemission signal from the sideband as shown in Fig.~\ref{fig:LAPE_trace}. The asymmetry in the time-trace arises from weak post-pulses in the pump pulse, which has been confirmed using a frequency-resolved optical gating (FROG) measurement.\cite{trebino2012frequency} The convolution of a Gaussian probe pulse and the measured pump pulse is fitted to the measured sideband intensity. From this a full-width at half maximum of 21$\pm$5~fs for the EUV probe is extracted. 

\section{\label{sec:limitationsandperspectives}Perspective and Limitations of TR-MM}

In summary, we have presented a novel setup combining the capabilities of a momentum microscope with a table-top HHG, and a versatile pump beamline, all operated at 1~MHz repetition rate. We demonstrate its capabilities for static band mapping, as well as for femtosecond time-resolved momentum microscopy with access to dynamics in the full Brillouin zone of a material.

TR-MM follows up on preceding experiments using TR-ARPES, where typically the photoelectron energy and in-plane momentum in one direction is measured. Momentum microscopes combined with time-of-flight analyzers thus offer new perspectives, because of their simultaneous measurement of the photoelectron energy and complete in-plane momentum. However, due to the novel photoelectron detection scheme, new challenges arise as well. In the following, we outline the limitations and perspectives of the setup.

\subsection{\label{sec:limitations}Limitations of TR-MM}

Limitations of the momentum microscopy have been reviewed by multiple groups.\cite{kromker_development_2008, schonhense_space-_2015, tusche_spin_2015, medjanik_direct_2017, Schonhense18njp, kutnyakhov_time-_2020, haag_time-resolved_2019, tusche_multi-mhz_2016, fedchenko_high-resolution_2019} Generally, the count rate and the signal-to-noise ratio of the MM are limited by the photoelectron detection scheme. Within the dead time of the delay line detector (DLD), it is not possible to detect more than one photoelectron, generally limiting the count rate to one photoelectron per laser pulse. It has been shown that segmented detectors composed of multiple DLDs can be used to increase the count rate.\cite{kutnyakhov_time-_2020} Alternatively, it is beneficial to increase the repetition rate of the excitation light source to improve the count rate. However, it is necessary to also consider aging of the MCP, which becomes increasingly relevant above $10^6$ counts per second for the used MCP in the DLD. The typical lifetime of a state-of-the-art MCP is $> 5000$ h at $10^{6}$ cps equally distributed. In terms of integral load on the MCP a doubling of the repetition rate, while maintaining the one electron per pulse regime, will thus lead to a halved lifetime. Additionally, pump-induced secondaries at small momenta can lead to an inhomogeneous degradation of the MCP, which might shorten the lifetime further. Therefore, running our experiment at 1 MHz repetition rate with slightly less than 1 photoemitted electron/pulse is a good trade off between maximum electron count rate and detector lifetime. 
Also, for instrument settings where all photoemitted electrons are detected, single-electron counting on the DLD naturally avoids space-charge effects, a common cause of signal distortion in photoemission spectroscopy.\cite{passlack_space_2006, Hellmann09prb, Oloff16jap, Schonhense18njp} 

The MM is designed such that each photoelectron leaving the surface region enters the electrostatic lens system. While this is desirable in order to measure the full width of the photoemission horizon, it as well intrinsically limits the signal-to-noise ratio if only a narrow ($E$, $k_x$, $k_y$)-region is of interest in the experiment; the maximum accessible $10^6$~cps are distributed over a larger energy and in-plane momentum region. The effective useful count rate can be improved by aligning the microscope such that only the desired momentum region is projected onto the MCP, as done for our measurement of the K-point of graphene in Fig.~\ref{fig:Occupied_Band_structure_Graphene}~(b). Complementary, the application of an energy barrier in the electrostatic lens system can block photoelectrons with low kinetic energies before they enter the time-of-flight detector.\cite{tusche_multi-mhz_2016} In these scenarios, however, space-charge effects may arise due to the increased photoelectron density at the sample and in the electrostatic lens system up to the filtering mechanism.

Signal distortions due to space-charge effects become even more serious when the pump pulses lead to significant photoemission. This is most-relevant with infrared to visible frequencies at higher pump fluences, where only a few photons are necessary to overcome the work function and lead to additional photoemitted electrons. Photoelectrons thus generated via the process of multi-photon photoemission,\cite{saathoff_laser-assisted_2008, Reutzel19prx} being possibly plasmonically enhanced,\cite{Merschdorf00apa, Reutzel19prl} contribute to the overall count rate, limiting the effectively useful count rate in the region of interest. Furthermore, electrons emitted via secondary scattering processes or by field emission due to the large extractor voltage can give strong background signals. Those photoelectrons are typically localized at small kinetic energies and in-plane momenta and might be suppressed by placing apertures or high-pass filters, as discussed above. 

Similarly, it is possible to select a different high harmonic to optimize the useful count rate. The MM collects all photoelectrons within the photoemission horizon, as defined by Eq.~\ref{eq:photoemission_horizon}. Thus, by reducing the EUV energy, the counts are distributed over a smaller energy-momentum volume, facilitating an increased signal-to-noise ratio for the desired region of interest. For example, when considering an imaginary sample with a ($E$, $k_x$, $k_y$)-independent flat density of states, non-dispersive bands in $k_\perp$-direction, and a $\hbar\omega$-independent dipole matrix element, photoexcitation with the 9\textsuperscript{th} harmonic compared to the 11\textsuperscript{th} harmonic would distribute the count rate over a 43$\%$ smaller overall ($E$, $k_x$, $k_y$)-volume due to a smaller photoemission horizon. Note of course, that still the EUV energy has to be chosen sufficiently high to reach the in-plane momentum of the spectroscopic feature. 

In pump-probe spectroscopy, it is commonly assumed that the probed system returns to the ground state well before the next pair of pump and probe pulses arrive to the sample. With increasing repetition rate of the experiment, it becomes more important to validate that assumption. In our experiments on graphene, the charge-carrier dynamics is reversible when driven at 1~MHz. However, we observe a significant heating of the sample due to the average incident power of the pump pulses used for the excitation of the sample. This is detected using a temperature sensor which is incorporated in the copper heatsink on which the sample is mounted, meaning that the measured 60\textcelsius{} increase is most likely an underestimate of the real temperature at the active surface area for our settings. Depending on the sample under investigation, it might therefore be necessary to reduce the pump-induced heating via a reduction of the experiment´s repetition rate at the cost of increased measurement time.

\subsection{\label{sec:perspectives}Perspectives of TR-MM}

The momentum microscope combined with a high-repetition rate EUV light source and a versatile pump beamline is a promising instrument to study the non-equilibrium dynamics of solid-state materials on the few-femtosecond to picosecond time scale.The simultaneous collection of ($E$, $k_x$, $k_y$)-resolved data stacks for each time step $\Delta t$ provides full information on the ultrafast non-equilibrium dynamics within one experiment. Vice-versa, the usage of hemispherical analyzers requires a stepwise collection of the full information on ($E$, $k_x$, $k_y$). However, with hemispherical analyzers, a better signal/noise ratio can be achieved for a reduced $(E,k_{x,y})$ cut and comparable integration times.

Furthermore, we would like to point out additional possibilities offered by the momentum microscope. As the MM offers a real-space mode similar to a photoemission electron microscope (PEEM), it is possible to place apertures in the real-space image. This enables the measurement of photoelectron spectroscopy data from microscopic regions on the surface.\cite{felter_momentum_2019} Such experiments are useful for samples that are only available in microscopic sizes, such as stacked (twisted) transition metal dichalcogenides or graphene.\cite{Mak10prl, Geisenhof19acsanm} Conversely, it is also possible to place apertures in the Fourier-plane in order to record real-space microscopy images at specific photoelecton momenta. This brings dark-field imaging capabilities to photoelectron microscopy. Additionally, the real-space imaging combined with the TOF allows energy-filtered detection of real-space images. In a pump-probe approach, such experiments might be used to observe charge transfer processes across interfaces in real-time.\cite{Man17natnano} 

Finally, the combination of the time-of-flight momentum microscope with an imaging spin-filter allows for efficient and simultaneous detection of multi-dimensional spin-resolved data sets.\cite{schonhense_space-_2015,SUGA2015119,SCHONHENSE201719} In an optical-pump~-~EUV-probe excitation scheme \cite{Plotzing2016,Eich:2017im, PhysRevLett.121.087206}, the simultaneous access to ultrafast energy-momentum-spin-resolved charge carrier dynamics is within reach.

\section{ACKNOWLEDGEMENTS}
M.K., C.M., D.S. and S.M. acknowledge support from the German Science Foundation through SFB 1073, project B07. G.S.M.J. and M.R. acknowledge funding by the Alexander von Humboldt Foundation. S.S. acknowledges the Dorothea Schl\"ozer Postdoctoral Programme for Women. The authors thank Klaus Pierz, Davood Momeni Pakdehi and Hans Werner Schumacher from PTB Braunschweig for providing the graphene sample.

\section{Data availability}
The data that support the findings of this study are available from the corresponding author upon reasonable request.
\section{REFERENCES}
\bibliography{paper_momi_setup}

\end{document}